\documentclass[ reprint,amsmath,amssymb,aps,superscriptaddress]{revtex4-1}
\usepackage{color}
\usepackage{amsmath}
\usepackage{graphicx}
\usepackage{dcolumn}
\usepackage{multirow}
\usepackage{bm}
\usepackage{textcomp}

\begin{document}
\preprint{APS/123-QED}

\title{Hacking Quantum Key Distribution via Injection Locking}%

\author{Xiao-Ling Pang}
\affiliation{Center for Integrated Quantum Information Technologies (IQIT), School of Physics and Astronomy and State Key Laboratory of Advanced Optical Communication Systems and Networks, Shanghai Jiao Tong University, Shanghai 200240, China}
\affiliation{CAS Center for Excellence and Synergetic Innovation Center in Quantum Information and Quantum Physics, University of Science and Technology of China, Hefei 230026, China}

\author{Ai-Lin Yang}
\affiliation{Center for Integrated Quantum Information Technologies (IQIT), School of Physics and Astronomy and State Key Laboratory of Advanced Optical Communication Systems and Networks, Shanghai Jiao Tong University, Shanghai 200240, China}
\affiliation{CAS Center for Excellence and Synergetic Innovation Center in Quantum Information and Quantum Physics, University of Science and Technology of China, Hefei 230026, China}

\author{Chao-Ni Zhang}
\affiliation{Center for Integrated Quantum Information Technologies (IQIT), School of Physics and Astronomy and State Key Laboratory of Advanced Optical Communication Systems and Networks, Shanghai Jiao Tong University, Shanghai 200240, China}
\affiliation{CAS Center for Excellence and Synergetic Innovation Center in Quantum Information and Quantum Physics, University of Science and Technology of China, Hefei 230026, China}

\author{Jian-Peng Dou}
\affiliation{Center for Integrated Quantum Information Technologies (IQIT), School of Physics and Astronomy and State Key Laboratory of Advanced Optical Communication Systems and Networks, Shanghai Jiao Tong University, Shanghai 200240, China}
\affiliation{CAS Center for Excellence and Synergetic Innovation Center in Quantum Information and Quantum Physics, University of Science and Technology of China, Hefei 230026, China}

\author{Hang Li}
\affiliation{Center for Integrated Quantum Information Technologies (IQIT), School of Physics and Astronomy and State Key Laboratory of Advanced Optical Communication Systems and Networks, Shanghai Jiao Tong University, Shanghai 200240, China}
\affiliation{CAS Center for Excellence and Synergetic Innovation Center in Quantum Information and Quantum Physics, University of Science and Technology of China, Hefei 230026, China}

\author{Jun Gao}
\affiliation{Center for Integrated Quantum Information Technologies (IQIT), School of Physics and Astronomy and State Key Laboratory of Advanced Optical Communication Systems and Networks, Shanghai Jiao Tong University, Shanghai 200240, China}
\affiliation{CAS Center for Excellence and Synergetic Innovation Center in Quantum Information and Quantum Physics, University of Science and Technology of China, Hefei 230026, China}

\author{Xian-Min Jin}
\thanks{xianmin.jin@sjtu.edu.cn}
\affiliation{Center for Integrated Quantum Information Technologies (IQIT), School of Physics and Astronomy and State Key Laboratory of Advanced Optical Communication Systems and Networks, Shanghai Jiao Tong University, Shanghai 200240, China}
\affiliation{CAS Center for Excellence and Synergetic Innovation Center in Quantum Information and Quantum Physics, University of Science and Technology of China, Hefei 230026, China}

\date{\today}

\pacs{Valid PACS appear here}

\begin{abstract}
Unconditionally secure communication, being pursued for thousands of years, however, hasn't been reached yet due to continuous competitions between encryption and hacking. Quantum key distribution (QKD), harnessing the quantum mechanical nature of superposition and non-cloning, may promise unconditional security by incorporating the one-time pad algorithm rigorously proved by Claude Shannon. Massive efforts have been made in building practical and commercial QKD systems, in particular, decoy states are employed to detect photon-number splitting attack against single-photon source loophole, and measurement-device-independent (MDI) QKD has further closed all loopholes in detection side, which leads to a seemingly real-life application. Here, we propose and experimentally demonstrate an MDI-QKD hacking strategy on the trusted source assumption by using injection locking technique. Eve injects near off-resonance photons in randomly chosen polarization into sender's laser, where injection locking in a shifted frequency can happen only when Eve's choice matches with sender's state. By setting a shifted window and switching the frequency of photons back afterwards, Eve in principle can obtain all the keys without terminating the real-time QKD. We observe the dynamics of a semiconductor laser with injected photons, and obtain a hacking success rate reaching 60.0\% of raw keys. Our results suggest that the spear-and-shield competitions on unconditional security may continue until all potential loopholes are discovered and closed ultimately.
\end{abstract}

\maketitle

\begin{figure*}
	\centering
	\includegraphics[width=2\columnwidth]{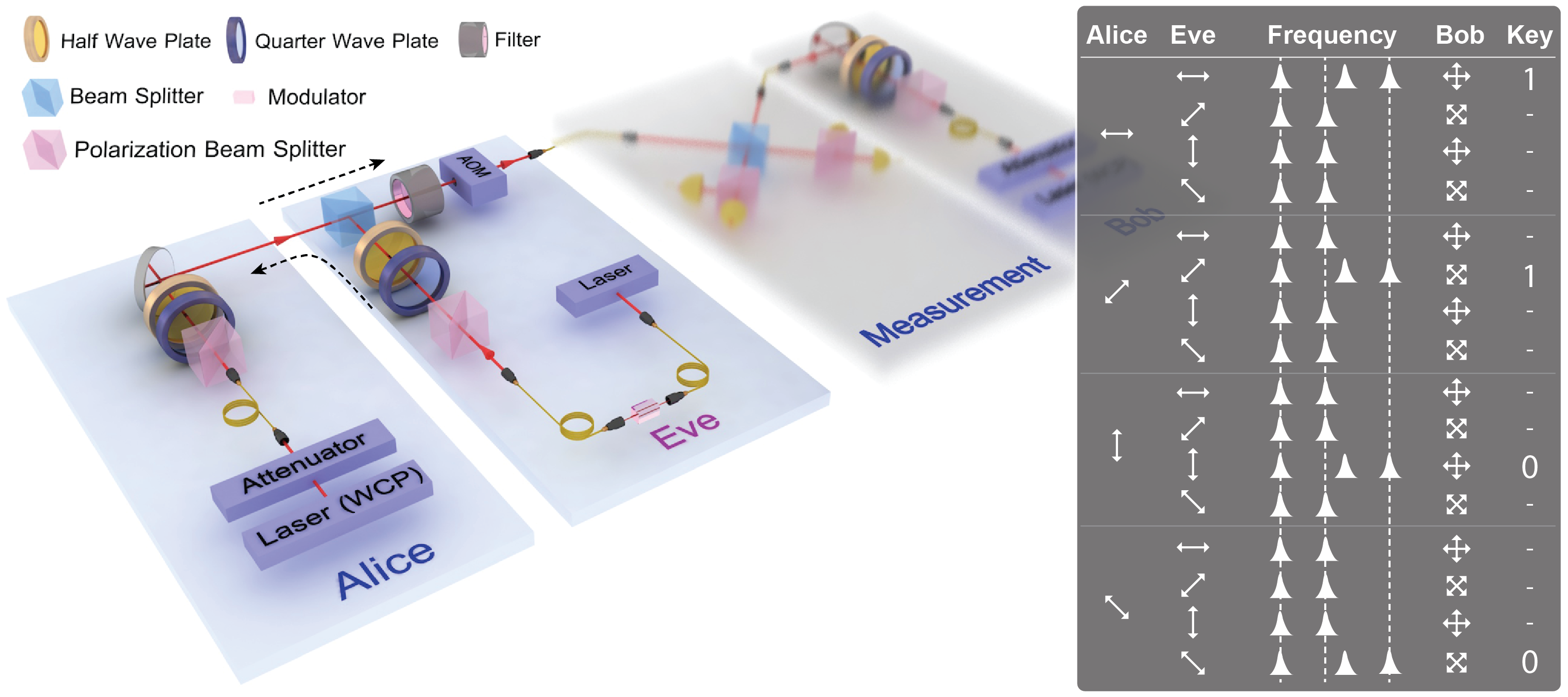}\\
	\caption{\textbf{Sketch of injection locking strategy hacking an MDI-QKD system.} At Alice's station, the WCP source is realized by a high photon-flux laser and isolators. The wavelength of Alice's laser is 852.355 nm, while that of Eve's laser is detuned by 251 MHz. Eve inserts a beamsplitter into the communication channel, and injects pulses with four randomly coded polarization states into Alice's site. A Fabry-Perot cavity serves as a spectrum filter to block unwanted photons, and an acoustic optical modulator (AOM) is utilized to shift frequency back to Alice's, and also works as a temporal filter. The complete set of polarization-matching situations of Alice and Eve is shown as the insert. For the matched cases,the emission frequency of Alice's laser is shifted well and become very distinct from the unmatched cases, and therefore can be selected by the cavity filter to form an effective information leakage. The dashed lines is for eye guiding to indicate the frequency of encoded photons from Alice at different stages.}
	\label{f1}
\end{figure*}

\section*{I. Introduction}

\noindent QKD is the best-known application of quantum cryptography, capable of distributing secure keys between two communication parties (known as Alice and Bob) in the presence of eavesdroppers (Eve). Both theoretical and experimental accomplishments have been made in the past decades \cite{Bennett_Cryptol_1992,Ursin_nphys_2007,Wang_nphoton_2013,Jiling_OE_2016}, and commercial QKD systems are now available on the market providing enhanced security for communication. Nevertheless, real-life devices are hard to conform with the hypotheses of theoretical security proofs, leading to continuous hacking events targeting at imperfections of certain devices in actual QKD implementations.

Enormous efforts have been made in developing hacking strategies and their countermeasures for improving the security of real-life QKD systems. Apart from few source-side attacks \cite{TangYL_PRA_2013,sun_PRA_2015,Lucamarini_PRX_2015}, most of the attacks target at the physical loopholes at detection side, which is the most vulnerable part in QKD setups, such as time shift \cite{Makarov_PRA_2006,Qi_QIC_2007,Zhao_PRA_2008}, time information \cite{Lamas_OE_2007}, detector control \cite{Lydersen_nphoton_2010,YuanZL_nphoton_2010,Gerhardt_ncommu_2011}, detector dead time \cite{Weier_NJP_2011} and channel calibration \cite{Jain_PRL_2011} attacks. A straightforward way to resist those attackers is to add corresponding countermeasures into QKD systems for each certain loophole \cite{YuanZLL_APL_2011,Honjo_OE_2013}. For instance, decoy-state protocol \cite{Hwang_PRL_2003,Wang_PRL_2005,Lo_PRL_2005}, marked as a milestone towards practical QKD, utilizes different intensities to detect photon-number splitting attack, allowing the same security level for weak coherent light sources as for perfect single-photon sources. 

DI-QKD \cite{Masanes_ncommu_2011,Reichardt_nature_2013,Vazirani_PRL_2014} is proposed to waive all the assumptions of trusted parties, either detection or source side, whereas it requires a detection efficiency higher than 80\%. While DI-QKD can close all the physical loopholes, its stringent requirement on detection efficiency is extremely challenging and remains to be met, limiting its application in ambient conditions. MDI-QKD is developed to close all potential detection-side loopholes, and most importantly, its implementation can be realized with current technology with acceptable key rate and channel distance \cite{Samuel_PRL_2012,Lo_PRL_2012}. MDI-QKD, together with decoy-state protocol, has closed almost all the known attacks on the remaining physical loopholes, and therefore provides a practically feasible and seemingly sufficient security.

Here, we show that it is still possible to attack a mature MDI-QKD system by conceiving a novel scheme and exploiting a new device loophole in source side. We demonstrate that Eve can control Alice's source by forcing her laser to be resonant at a designed frequency. We analyze and experimentally observe the injection-locking dynamics of a semiconductor laser, and obtain a hacking success rate (overlapped keys between Eve and Bob) of 60.0\% of raw keys and error rate of 6.1\%, with an appropriate injection power 110 nW. To test the robustness of hacking scheme in practice, a series of countermeasures made by Alice are conducted by adding 1 dB and 3 dB isolation into the channel, which leads to the slight declines of success rate to 44.0\% and 36.1\% respectively, but there is still considerable information leakage. Furthermore, Alice's isolation can be eliminated by increasing the injection laser power. We also calculate the secure key rates (SKRs) and quantum bit error rates (QBERs) as functions of channel length under different injection power, and we discuss the effect of injection power on hacking success rate. Given the assumption that Eve has unlimited capabilities as long as being subject to quantum mechanics, for example infinite laser power, she can always hack MDI-QKD systems with constantly high success rate.

\begin{table}[!b]
	\centering	
	\caption{\textbf{Standard post-selection process in QKD and Eve's strategy.} After the measurement-side announces successful results, Alice and Bob post-select the incidents that they use the same bases. Eve discards the events that her bases are different from Alice's, and keeps the rest. R: rectilinear basis; D: diagonal basis.}	
	\label{tab1}
	\begin{tabular}{p{1.2cm}<{\centering} p{0.9cm}<{\centering} p{1cm}<{\centering} p{2.4cm}<{\centering}  p{2.4cm}<{\centering}}
		\hline\noalign{\vskip 0.6mm}
		\hline\noalign{\vskip 0.2cm}		
		\textbf{Alice} &  \textbf{Bob} & \textbf{Eve} & \textbf{Post-Selection} & \textbf{Eve's Strategy} \\[0.10cm]		
		\hline\noalign{\vskip 0.11cm}		
		R & R &	R & \multirow{2}*{Keep} & \multirow{2}*{Keep} \\[0.00cm]
		D & D &	D \\[0.00cm]
		\hline\noalign{\vskip 0.6mm}
		R & R &	D & \multirow{2}*{Keep} & \multirow{2}*{Discard} \\[0.00cm]
		D & D &	R \\[0.00cm]
		\hline\noalign{\vskip 0.6mm}
		R & D &	R/D & \multirow{2}*{---} & \multirow{2}*{---} \\[0.00cm]
		D & R &	R/D \\[0.00cm]
		\hline\noalign{\vskip 0.6mm}
		\hline\noalign{\vskip 0.6mm}
	\end{tabular}
	\par
\end{table}

\section*{II. HACKING STRATEGY} 

As is shown in FIG. 1, Alice and Bob prepare weak coherent pulses (WCPs) in four BB84 polarization states, and send them to an untrusted relay to perform a Bell state measurement as a typically MDI-QKD does \cite{Lo_PRL_2012}. To demonstrate the hacking strategy, Eve (known as master laser) injects a beam of near off-resonance pulse with four randomly chosen polarization states into Alice's laser (known as slave laser) backward. The slave laser contains a 30-dB built-in isolator. From the inset of Figure 1, we can see that injection locking in a shifted frequency can happen only when Eve's choice matches with Alice's choice. By setting a shifted window via a frequency filter, Eve maximizes the transmittance for the photons in the matched cases. Under this circumstance, Eve in principle can obtain all the keys. It should be noticed that the frequency of the transmitted photons can be converted back, before interfering with Bob's photons, thus the whole hacking process doesn't terminate the real-time QKD.

\begin{figure}
	\includegraphics[width=1\columnwidth]{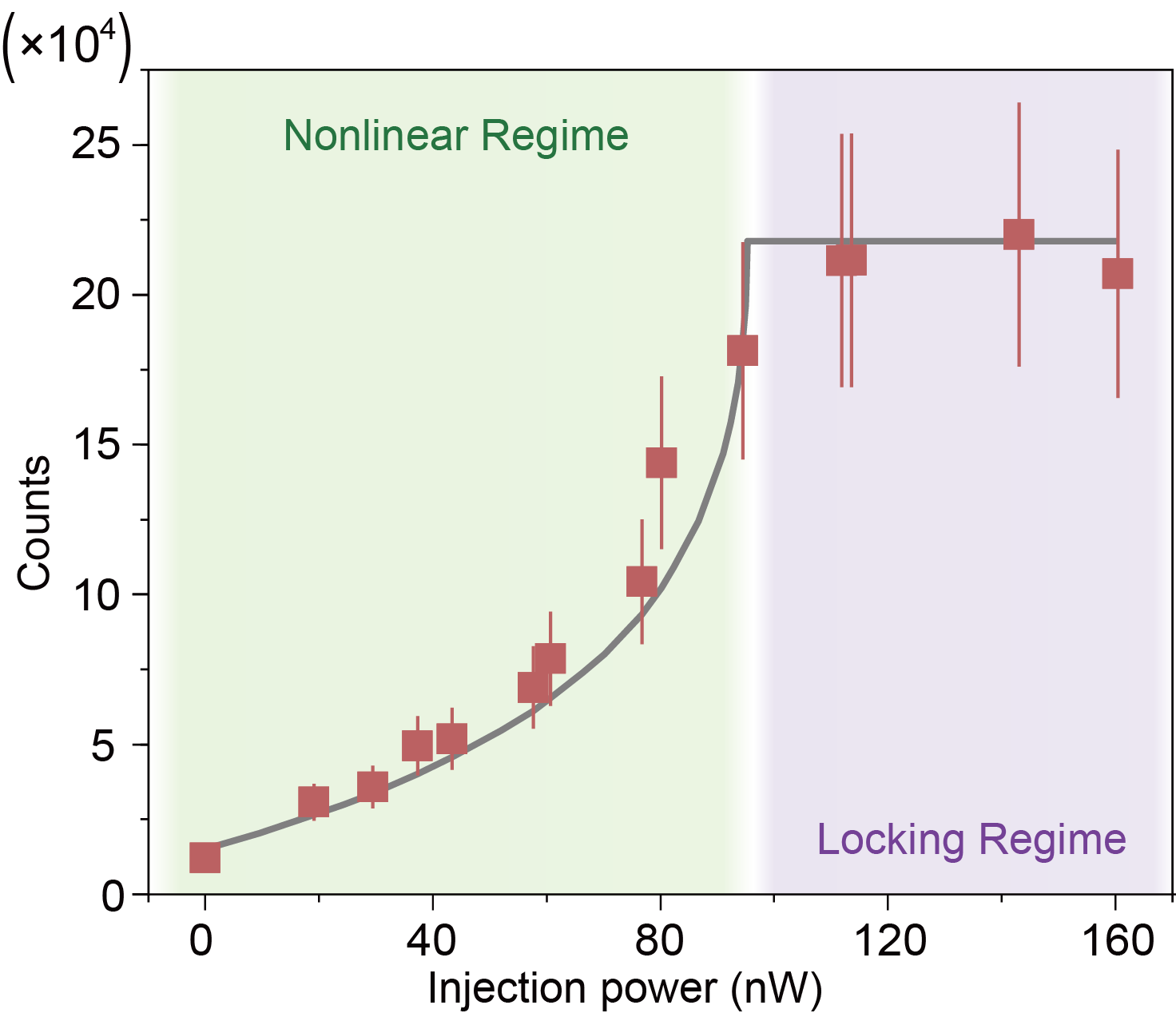}\\
	\caption{\textbf{Theoretical and experimental results of the dynamical behavior of an injected semiconductor laser}. Dark gray curve is the theoretical value of counts, and red squares are the measured counts with the increase of injection power. Area in green represents nonlinear process, where multi-wave mixing appears. Area in purple represents stable locking process. Error bar comes form transmittance jitter of the cavity filter in a slightly disturbed environment.}
	\label{f2}
\end{figure}

The detuned pulses from Eve's master laser are prepared in four polarization states randomly. Alice's slave laser (a semiconductor laser in a wavelength of 852.355 nm near cesium D2 line) can be in the stage of either stable locking or nonlinear process, when it is externally injected with monochromatic photons under various injection-power levels \cite{Adler_PIRE_1946}. In the nonlinear regime, many frequency components are generated, and multimode competition appears. With the increase of injected photons, the frequency of the slave laser approaches that of master's completely when the injection power exceeds the stable locking threshold. A Fabry-Perot cavity that only resonant with the frequency of the master laser can serve as a spectrum filter to block unlocked photons and pick out the matched signals. The following acoustic optical modulator can serve as frequency shifter as well as a temporal filter, which further contributes to block unwanted photons. See Appendix for more details.

A hacking event is considered to be successful if Eve chooses the same bases with Alice: let the same polarized photons pass through the cavity filter, while block orthogonally polarized photons. Afterwards, Eve monitors Alice and Bob's post-selection results, as in BB84 protocol, and keeps the data where she uses the same bases with Alice, and discards the rest according to TABLE I. Moreover, once a quantum communication process is completed, Eve applies the same operation with Alice according to the announced measurement results (typically see TABLE I in reference \cite{Lo_PRL_2012}), to guarantee that she possesses the right bit streams as well as Alice's and Bob's.

\section*{III. EXPERIMENTAL RESULTS }

In order to increase hacking success rate and decrease error rate, the injection power must be strong enough to lock slave laser, and simultaneously half of this power should be too weak to do so. As is shown in FIG. 2, we simulate the dynamics of a single-mode semiconductor laser with injected photons theoretically, and obtain the intensity of output photons by numerical simulation and spectrum analysis, see Appendix for more details. We further experimentally demonstrate this dynamics process by setting a series of injection power and recording counts of output photons after the cavity filter, as the square points presented in FIG. 2, which follow our theoretical simulation very well. With such a test, we can clearly see the nonlinear regime and locking regime, and more importantly it identifies a transition point. The pulses that match with Alice's polarization should have enough power to enable injection locking, and all other pulses are in the nonlinear regime. Apparently, the sharper the transition is, the higher success rate and lower error rate we can achieve in the hacking process. 

\begin{figure}
	\centering
	\includegraphics[width=1\columnwidth]{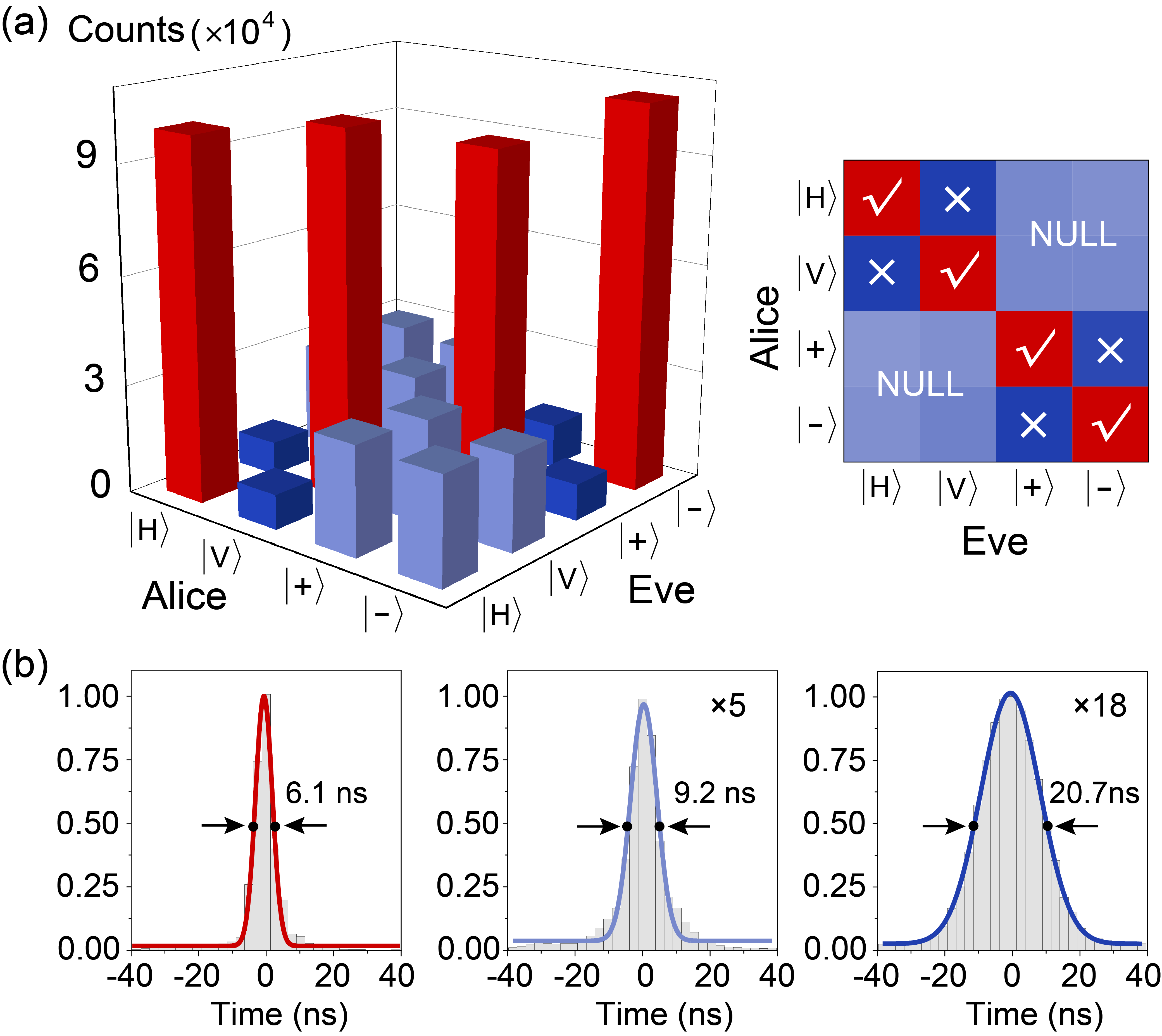}\\
	\caption{\textbf{Experimental results of a typical hacking performance.} \textbf{a.} Photon number distribution of all 16 combinations of Eve's and Alice's polarization states. Columns in red ($\surd$) are the successful-hacked keys. Columns in dark blue ($\rm{\times}$) are the observed errors in the keys. Columns in light blue (NULL) are invalid events that will be discarded by Eve unilaterally after a standard post-selection in QKD. \textbf{b.} Shape of pulses that transmitted through the cavity filter when the polarization angles between Eve and Alice are in $0^\circ$ (left), $45^\circ$ (middle) and $90^\circ$ (right). Arrows refer to FWHM fitted with Gaussian function. The data is obtained under the condition of the injection power of 110 nW after a 30-dB built-in isolator.}
	\label{f3}
\end{figure}

\begin{figure*}
	\centering
	\includegraphics[width=1.8\columnwidth]{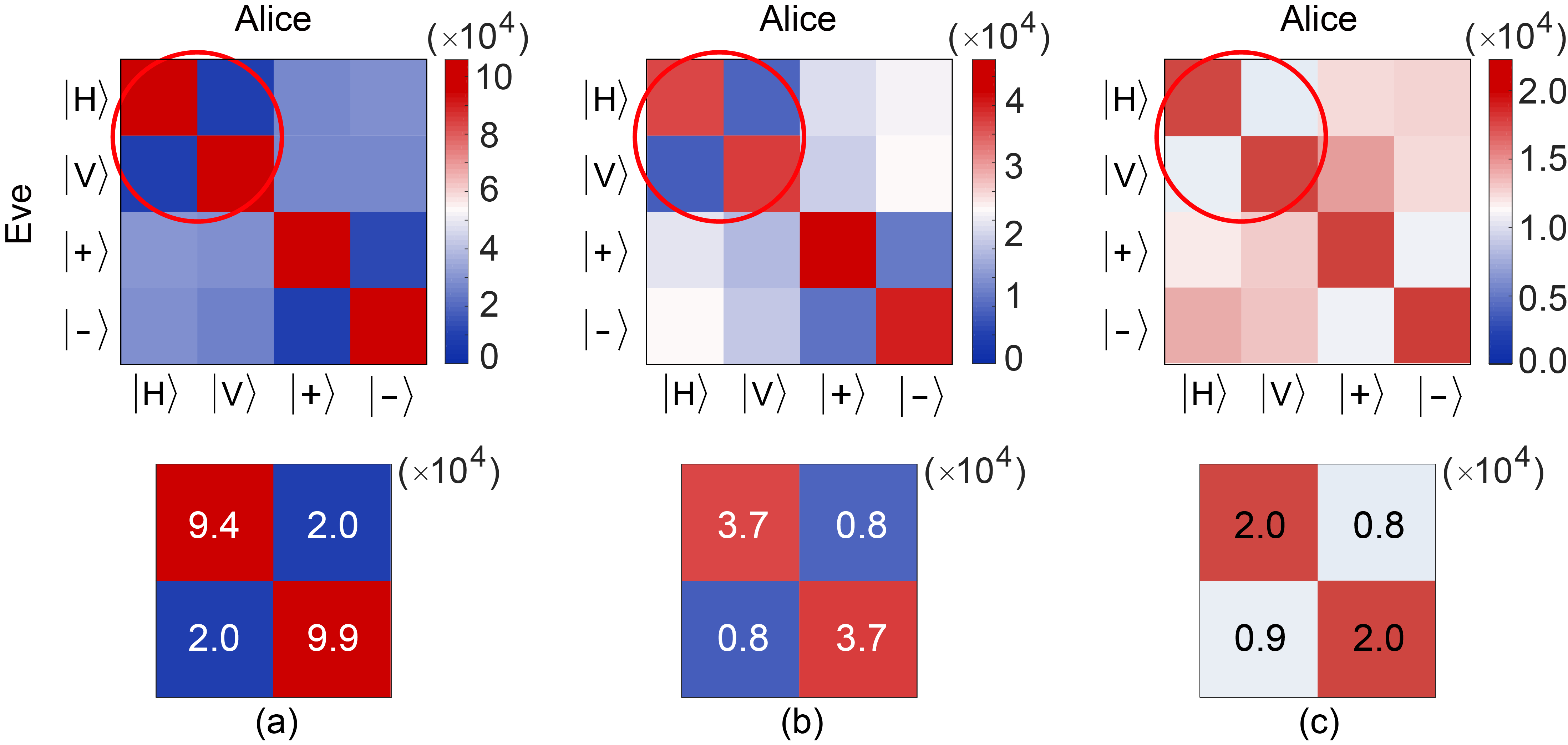}\\
	\caption{\textbf{Hacking performance under the isolation countermeasure strategy with additional channel isolation.} Eve's injection power is fixed at 110 nW, then counts of transmitted photons are measured under the condition of adding (\textbf{a}) 0 dB, (\textbf{b}) 1 dB and (\textbf{c}) 3 dB isolation. Figures above present the photon number distribution of all 16 combinations of Eve's and Alice's polarization states. The insets below emphasize on error rate in successful-hacked events, as circled in red, and the values in boxes detail the counts of photons under corresponding situations. Apparently,  still considerable successful-hacked keys can be obtained in all the cases.} 
	\label{f4}
\end{figure*}

\begin{table*}
	\centering
	\caption{\textbf{Experimental results of success rate and error rate of hacking strategy under different situations.} }
	\label{tab2}
	\begin{tabular}{
			p{4.4cm}<{\centering}  p{1.3cm}<{\centering} 
			p{1.3cm}<{\centering}  p{1.3cm}<{\centering}
			p{0.5cm}<{\centering}  p{1.1cm}<{\centering}
			p{0.5cm}<{\centering}  p{3.2cm}<{\centering}}
		\hline\noalign{\vskip 0.6mm}
		\hline\noalign{\vskip 0.14cm}
		
		\textbf{Light Source} &   & \multicolumn{2}{c}{\textbf{WCP}}  &   &   &   &\textbf{ High photon flux}  \\[0.06cm]
		\cline{2-6} \cline{8-8} \noalign{\vskip 0.8mm}	
		
		\textbf{Injection power (nW)} &   & \textbf{110} &   &   & \textbf{119} &  & \textbf{110}  \\[0.08cm]
		\cline{2-4} \cline{6-6} \cline{8-8} \noalign{\vskip 0.8mm}		
		
		\textbf{Isolation (dB)} & \textbf{0}  & \textbf{1} & \textbf{3} &   & \textbf{0} &  & \textbf{0}  \\[0.08cm]
		\hline\noalign{\vskip 0.8mm}
		
		\textbf{Success rate (\%)} & 60.0 & 44.0 & 36.1 &   & 47.0 &  & 67.4  \\[0.08cm]
		\textbf{Error rate (\%)} & 6.1 & 10.0 & 15.4 &   & 9.8 &  & 1.7  \\[0.08cm]
		
		\hline\noalign{\vskip 0.6mm}
		\hline\noalign{\vskip 0.14cm}		
	\end{tabular}
\end{table*}

Based on the characterized dynamical behavior of Alice's laser shown in FIG. 2, we set the injection power at 110 nW, slightly stronger than the transition point to demonstrate a typical hacking performance. The measured photon flux transmitted through the cavity filter are shown in a 3-D histogram and its top view (see FIG. 3a and the inset), with all possible combinations of Eve's and Alice's polarization states. Light blue columns represent transmitted photons in the case of that Alice and Eve choose the different bases, and therefore Eve will discard these parts for removing eavesdropping failure after a standard post-selection in QKD. Red and dark blue columns represent transmitted photons when Alice and Eve choose the same bases, which are the keys to be kept. In particular, the red parts are correct-hacked keys, and dark blue parts represents errors.

Defining the eavesdropping success rate as the proportion of the overlapped keys between Eve and Bob in all events; and the eavesdropping error rate as the proportion of mistakes in kept data. The successful rate measured at the injection power of 110 nW of raw keys reaches 60.0\%, and the error rate is 6.1\%, and the remaining 33.9\% is the loss rate. The error mainly comes from the failure of blocking unwanted photons, due to the fact that the injection locking cannot be ceased completely in the nonlinear regime. Few errors come from the fluorescent noise of the slave laser, the scattering on the beam splitter, and also the limited extinction rate of the cavity filter. Pulse shapes of transmitted photons in three injection levels are illustrated as FIG. 3b. We can clearly see that the intensity obtained in locking regime is much stronger than that in nonlinear regime. 

\begin{figure*}
	\centering
	\includegraphics[width=2.06\columnwidth]{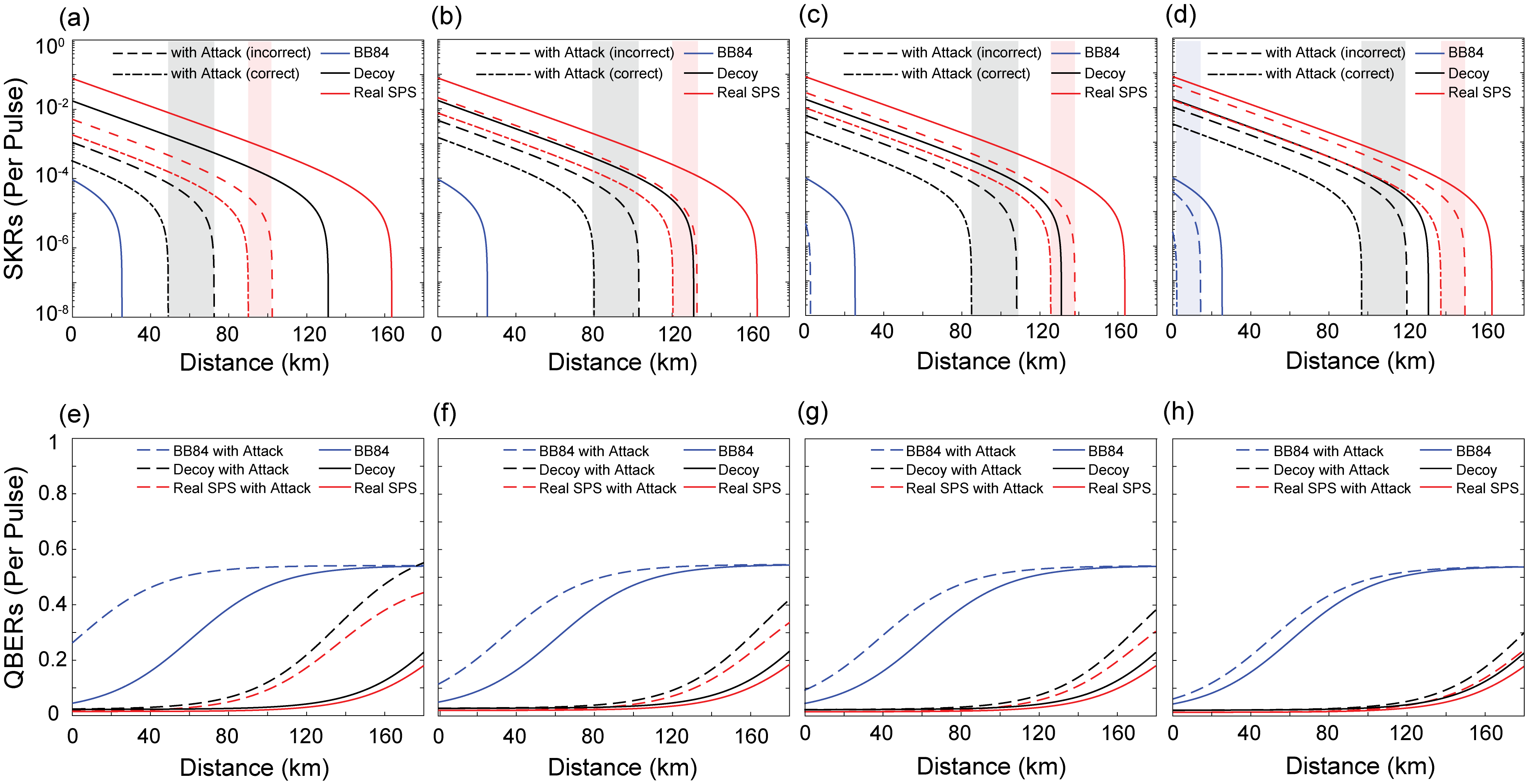}\\
	\caption{\textbf{SKRs and QBERs as functions of transmission distance.} Theoretical SKRs before and after injection locking attack are presented with the power of: \textbf{(a)} 50 nW, \textbf{(b)} 100 nW, \textbf{(c)} 150 nW and \textbf{(d)} 200 nW. The shaded areas mark the insecure communication ranges with different QKD implementations. Corresponding QBERs are shown below in \textbf{(e)-(h)}. Dashed (Solid) lines are values with (without) attack. Lines in red represent for real single photon sources (SPSs) cases, in black represent for decoy-state BB84 protocols, and in blue represent for original BB84 protocols.} 
	\label{f5}
\end{figure*}

\section*{IV. COUNTERMEASURES}

A very straightforward countermeasure strategy that we can think of against the attack is to add more isolation to block Eve's injection. We realize this by adding 1 dB and 3 dB isolation in Alice's station. As is shown in FIG. 4 and also in TABLE II, we do observe a decline of the hacking success rate of raw keys slightly from 60.0\% to 44.0\% and 36.1\%, and an increase of the hacking error rate from 6.1\% to 10.0\% and 15.4\%, respectively. Keeping other conditions unchanged, we can see that while the isolation countermeasure strategy can lower the hacking success rate and raise the error rate, unfortunately, still considerable part of keys is hacked successfully. Furthermore, Alice's isolation can be eliminated by increasing the injection laser power, which leads to a new transition point. Also, there are researches proving that the attenuation of some types of attenuator can be permanently decreased after laser damage \cite{Huang_arxiv_2019_att}. Given the assumption that Eve has unlimited capabilities as long as being subject to quantum mechanics, for example infinite laser power, she can always hack MDI-QKD systems with constantly high success rate.

More experimental results under different conditions are summarized in Table II. Hacking performance of WCP sources under four different injection power situations indicates the optimal hacking power and the robustness of the hacking strategy. We also demonstrate the additional experiment for hacking high photon flux source. Under this circumstance, the signal-to-noise ratio is increased by raising the signal and keeping the noise constant. The outcome reveals the negative influence of noise in our implementations, which can be improved by further reducing the scattering on the beam splitter and narrowing the window of the temporal filter. 

\begin{figure}
	\centering
	\includegraphics[width=0.9\columnwidth]{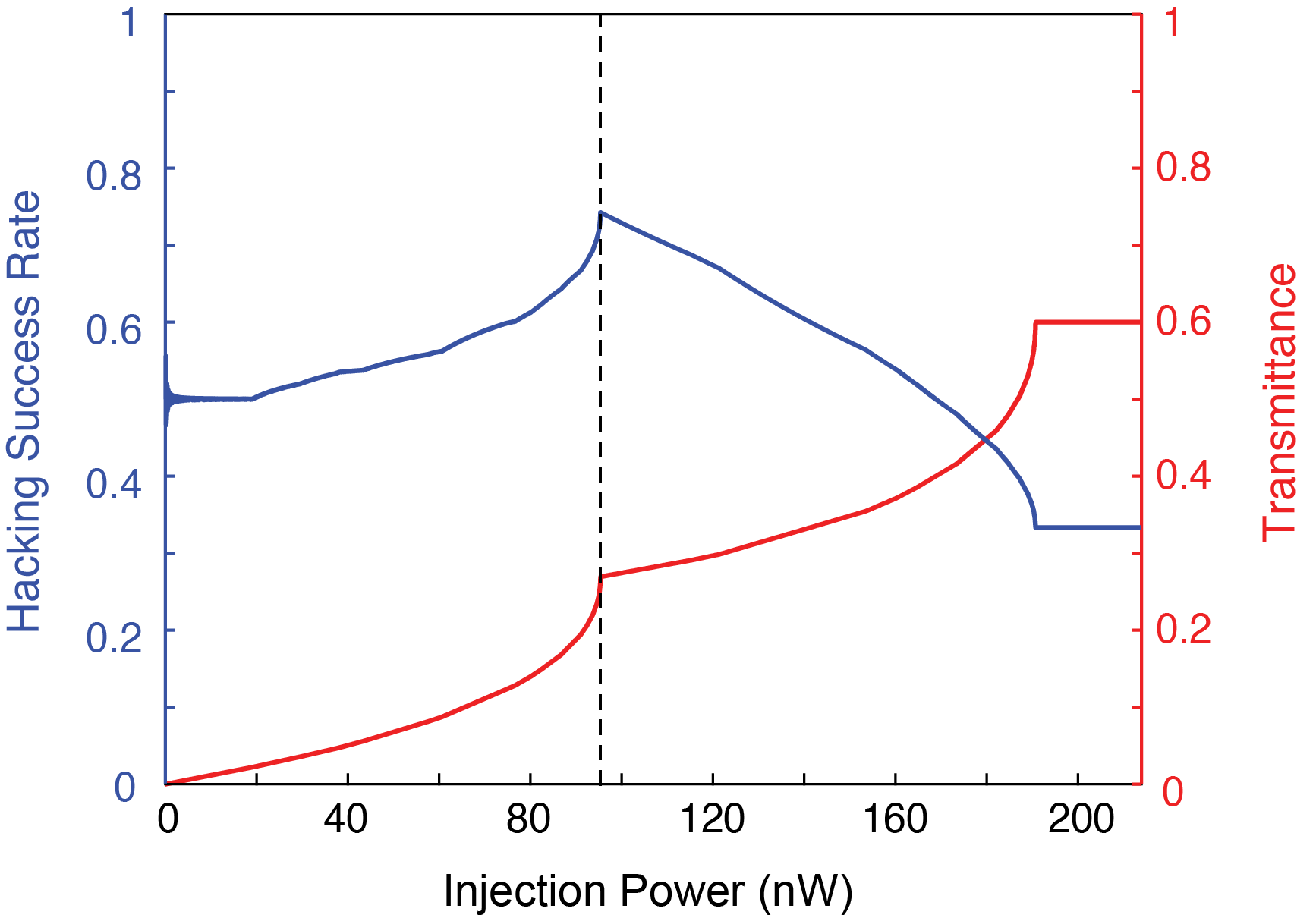}\\
	\caption{\textbf{Theoretical results of hacking success rate and transmittance as functions of injection power.} The blue line represents for the calculated hacking success rate, and the red line represents for the transmittance induced by attack. The dashed line in the middle indicates the optimal injection power with a maximal hacking success rate.}
	\label{f6}
\end{figure}

 \section*{V. EFFECT ON THE SECURITY OF QKD}
To analyze the security of keys, we calculate the SKRs and QBERs before and after the injection-locking attack (with 50, 100, 150 and 200 nW injection power respectively), as is shown in FIG. 5. The shaded areas mark the insecure communication ranges, in which the correct SKRs are positive while the incorrect SKRs are negative. In a secure communication, the calculated SKR is positive (Shannon mutual information $I(A,B)>I(A,E)$), and after error correction and privacy amplification, $I(A,E)$ could be small enough to be ignored, leading to secure keys. However, in the marked areas, the correct SKRs are negative ($I(A,B)<I(A,E)$), which means the communication is insecure that even after error correction and privacy amplification, $I(A,E)$ cannot be ignored. Therefore, Eve can eavesdrop the information of keys without being detected in the marked areas.

The influence of attack on the SKSs and QBERs becomes negligible with the injection power increases, which is mainly resulted from the increased channel transmittance, and this effect can be weakened by utilizing low-loss channel. Furthermore, in order to analyze the influence of injection power, we present the relationship between the injection power and the hacking success rate, as well as the relationship between the injection power and the channel transmittance, shown in FIG. 6. The hacking success rate increases first and then decreases, with a maximum value at around 100 nW injection power. However, still considerable numbers of keys up to 33\% have been eavesdropped by Eve over the entire region. It should be noticed that the optimal value for different QKD transmitters can be identified by Eve by scanning the injection power while checking the success rate in the post-selection process. Details and parameters in simulation are given in Appendix.

Another possible countermeasure strategy against source-side attack is to place classical detectors to monitor strong light from outside. But it has been proven that the current implementations of the monitoring detectors are incapable of being perfectly secure \cite{Shihan_PRA_2015}, and also it has the problem of sensitivity limitations. Meanwhile the whole system will have to become more complicated and expensive. So, we still need to be cautious about isolators and any other source-side loopholes, and to conduct more in-depth researches to make the practical QKD systems approaching a sufficient security.

\section*{VI. CONCLUSIONS}

In summary, we propose and demonstrate an MDI-QKD hacking strategy on the trusted source assumption by using injection locking technique, with a success rate of raw keys approaching 60.0\% and an error rate as low as 6.1\%. We also propose and demonstrate an isolation countermeasure strategy to test the robustness of the injection locking strategy. While the performance of the hacking strategy degrades, unfortunately, still significant amounts of keys are eavesdropped without being detected. Even worse, unconditional security of QKD is under the assumption that Eve has the power only subjected to quantum mechanics, apparently with infinite laser power, Eve will always be able to hack MDI-QKD systems even with the isolation protection.

After Eve has forced Alice to produce polarization states of different spectra, this has created a state-distinguishability side-channel that can be exploited in different attacks (not just the strategy demonstrated above). For example, Eve can do an intercept-resend attack, or one can use the spectra measured in a security proof and possibly obtain a positive key generation rate even in the presence of attack, as long as the attack is limited by isolators and the spectra are not too different between the states.

A sufficient security is possible if Alice adds many orders more isolation, since Eve cannot possess infinite laser power in practice. Of course, the whole system of  MDI-QKD will have to become more complicated and expensive. However, the main message we would like to deliver here is that there may exist many other physical loopholes remaining undiscovered when we believe that MDI-QKD has already been a very mature and commercially available solution. Our results suggest that the spear-and-shield competitions on unconditionally secure communication may continue, and hacking strategies, though their anti-hacking solutions may be readily available, should be enthusiastically pursued to discover and close all potential loopholes for genuinely unconditional security.

While this paper was under review, we learnt another laser seeding experiment that changes the intensity \cite{Huang_arxiv_2019} of Alice's laser rather than its wavelength.

\section*{ACKNOWLEDGMENTS}
This research was supported by the National Key R\&D Program of China (2019YFA0308700, 2017YFA0303700), the National Natural Science Foundation of China (11761141014, 61734005, 11690033), the Science and Technology Commission of Shanghai Municipality (STCSM) (17JC1400403), and the Shanghai Municipal Education Commission (SMEC) (2017-01-07-00-02- E00049). X.-M.J. acknowledges additional support from a Shanghai talent program.


\begin{table*}[t]
	\centering	
	\caption{\textbf{Parameters of semiconductor laser used in calculation.} }	
	\label{tab1}
	\begin{tabular}{p{2cm}<{\centering} p{3cm}<{\centering} p{2cm}<{\centering} p{6.2cm}<{\centering} }
		\hline\noalign{\vskip 0.6mm}
		\hline\noalign{\vskip 0.2cm}		
		$\alpha$	&	$4.5$	&	$m^{-1}$	&	Linewidth Enhancement Factor	 \\[0.10cm]		
		\hline\noalign{\vskip 0.11cm}		
		$G_N$ & $5 \times{10^3}$ & $s^{-1}$ & Gain Coefficient	 \\[0.10cm]		
		\hline\noalign{\vskip 0.6mm}
		$\kappa$ & $1.2136 \times{10^{11}}$ & $ps^{-1}$ & Fees-in Rate	\\[0.10cm]		
		\hline\noalign{\vskip 0.6mm}
		$T_S$ & $0.8059$ & $ns$ & Electro-hole Recombination Time	\\[0.10cm]		
		\hline\noalign{\vskip 0.6mm}
		$T_P$ & $1.1628$ & $ps$ & Photon Lifetime	\\[0.10cm]		
		\hline\noalign{\vskip 0.6mm}
		$P_0$ & $3.5183 \times{10^{5}}$ & $-$ & Original Photon Number	\\[0.10cm]		
		\hline\noalign{\vskip 0.6mm}
		$\alpha_m$ & $3896$ & $-$ & Facet Loss	\\[0.10cm]		
		\hline\noalign{\vskip 0.6mm}
		$\nu_g$ & $7.5\times 10^7 $ & $m/s$ & Velocity	\\[0.10cm]		
		\hline\noalign{\vskip 0.6mm}
		$\hbar\omega$ & $1.456$ & $eV$ & Photon Energy	\\[0.10cm]		
		\hline\noalign{\vskip 0.6mm}
		\hline\noalign{\vskip 0.6mm}
	\end{tabular}
	\par
\end{table*}

\section*{APPENDIX A: DETAILS OF THE SOURCE-SIDE ATTACK}

Here we present the detailed information of the hacking strategy via injection locking. The pulse width of Eve’s laser is around 10 ns, and the repetition rate is 10 MHz. Injection locking process in a semiconductor laser (known as the slaver laser) happens when it is externally injected with proper intensity of light (from the master laser). When the intensity of injected light approaches the threshold intensity of stable locking, the frequency of slaver laser will be resonant with that of master laser's  (see next section for dynamics of injection locking process). In order to obtain the injection locking threshold experimentally, we scan the intensity of injected light, and measure the counts of photons that transmits through the cavity filter. Results are shown in FIG. 2 of the main text.

\begin{figure}[t]
	\centering
	\includegraphics[width=1\columnwidth]{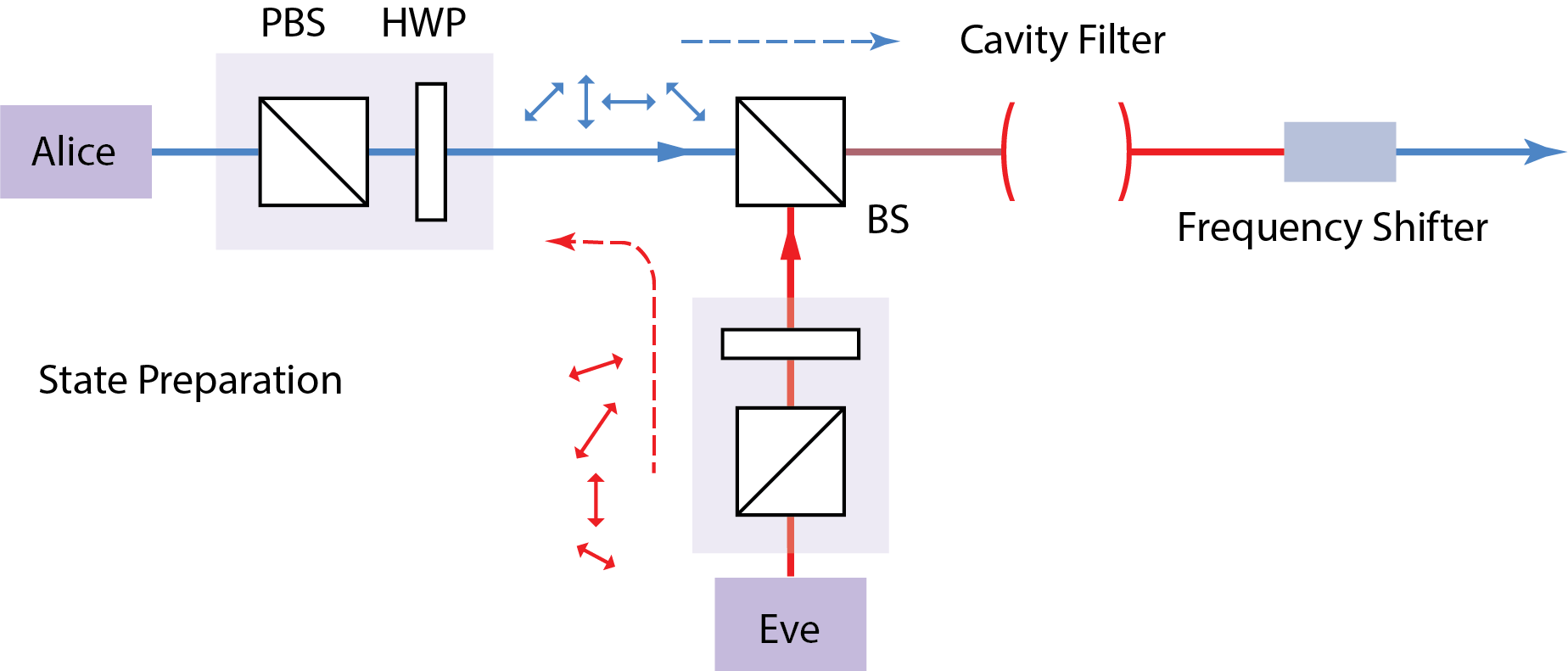}\\
	\caption{\textbf{Injection locking strategy of source-side attack.} Lines in blue (red) are photons with the frequency of the slave (master) laser. PBS: polarization beam splitter; BS: beam splitter; HWP: half wave plate.}
	\label{f1}
\end{figure}

In our hacking strategy, shown in FIG. 7: Alice prepares light pulses in four BB84 polarization states with a polarization beam splitter (PBS) and a half wave plate (HWP), and sends them to the measurement devices. As we know, Eve also prepares four randomly chosen polarization states, and injected to Alice's side via a beam splitter (BS). According to FIG. 7, there might be three situations:

(1) When the polarization states prepared by Eve and Alice are the same, all the photons from Eve's side can be injected into Alice's laser. Thus, the slaver laser is locked and the emission frequency is shifted to master laser's.

(2) When the polarization states prepared by Eve and Alice have a difference of 45 degree, half of the photons from Eve's side can be successfully injected into Alice's, and nonlinear process happens while the emission frequency is partially shifted to master laser's.

(3) When the polarization states prepared by Eve and Alice are perpendicular, none of the photons from Eve's side can be injected into Alice's, injection locking will not happen.

Therefore, we conclude that the attack works perfectly when the polarization prepared by Alice and Eve are the same. On the other hand, such hacking strategy increases the channel loss, induced by the photons in the frequency of Alice's laser which are then blocked by the cavity filter.

\begin{equation}\label{eq01}
 \begin{aligned}
\frac{d}{{dt}}{E_0}(t) = \frac{1}{2}\left( {1 + i\alpha } \right){G_N}\Delta N(t)E(t) + \kappa {E_x}{e^{i\nu t}}
 \end{aligned}
\end{equation}

\begin{equation}\label{eq02}
 \begin{aligned}
\frac{d}{{dt}}\Delta N(t) =  & - \left( {\frac{1}{{{T_S}}} + {G_N}{P_0}} \right)\Delta N(t) \\ 
					& - \left[ {\frac{1}{{{T_P}}} + {G_N}\Delta N(t)} \right]\left( {{{\left| {E(t)} \right|}^2} - {P_0}} \right)
 \end{aligned}
\end{equation}
where $E(t)$ represents the total changeable electric field; $E_x$ is the amplitude of injected photons, and $\nu$ is the difference of frequency between injected and original photons. Note that all electric fields mentioned here are normalized so that the square of their amplitudes give the corresponding photon numbers in laser cavity. $\delta N(t)$ represents the difference of carrier numbers between actual and original situations; $\kappa$ is the fees-in rate (reciprocal value of round-trip-time in laser cavity); $P_0$ stands for the number of original photons. Others are the parameters of the semiconductor laser, as is shown in TABLE III. 

\begin{figure*}[htb!]
	\centering
	\includegraphics[width=1.6\columnwidth]{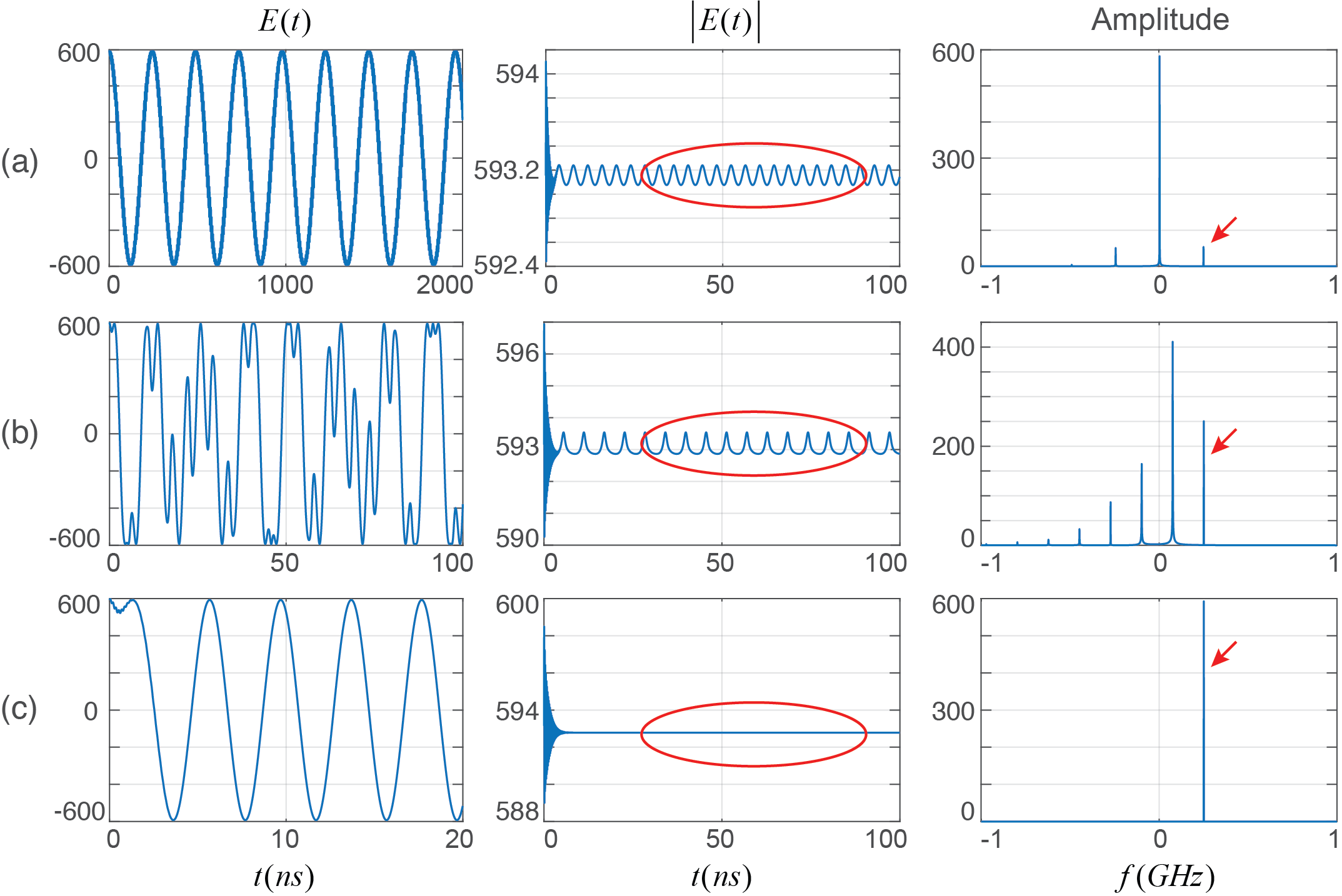}\\
	\caption{\textbf{Simulation results of injection-locking dynamics.} The varying electric field (left), amplitude (middle) and spectrum (right) results with \textbf{(a)} $E_x=0.3$, \textbf{(b)} $E_x=1.2$, \textbf{(a)} $E_x=1.7$. Red circles mark the regions used for Fourier analysis, and red arrows indicate the target frequency component.}
	\label{f1}
\end{figure*}

\section*{APPENDIX B: DYNAMICS OF A SEMICONDUCTOR LASER WITH INJECTION PHOTONS}

Here we present theoretical details of the dynamics of a single-mode semiconductor laser with external injection. When the slave laser injected with increasingly stronger external photon flux, it transforms from nonlinear regime to locking regime under weak injection situations, and after a complex intermediate region, it ends at a strong injection locking regime \cite{Annovazzi_IEEE_1994}. Here we just simulate weak injection situations needed in our experiment.

The external injection process is described by adding a forcing term to the Lang-Kobayashi model \cite{Mogensen_IEEE_1985}:

When injected with external photon flux, the variance of electric filed can be simulated by numerically integrating equations (1) and (2), and after applying Fourier transform, the amplitude of injection-amplified laser is calculated. Since what we concerned here is the power of the laser with a certain frequency, the relation between output power ($P_{power}$) and photon number in laser cavity ($P$) is described as \cite{Van_IEEE_1985}:

\begin{equation}\label{eq03}
 \begin{aligned}
{P_{power}} = \frac{1}{2}{\alpha _m}{\nu _g}\hbar \omega P
 \end{aligned}
\end{equation}

FIG. 8 demonstrates the varying electric field, amplitude and spectrum results with $E_x=$0.3 (a), 1.2 (b), 1.7 (c). The initial state in simulation is set to the original state of the slave laser, and thus it takes several nanoseconds to evolve. So we use the later part for Fourier analysis, which is circled in FIG. 8.

\begin{table*}[t]
	\centering	
	\caption{\textbf{Parameters used in simulation for calculating SKRs and QBERs.} }	
	\label{tab1}
	\begin{tabular}{p{1.5cm}<{\centering} p{2.5cm}<{\centering} p{1.5cm}<{\centering} p{11cm}<{\centering} }
		\hline\noalign{\vskip 0.6mm}
		\hline\noalign{\vskip 0.2cm}		
		$\alpha$	&	$0.2$	&	$dB/km$	&	Channel loss	 \\[0.10cm]		
		\hline\noalign{\vskip 0.11cm}		
		$Y_0$ & $2.6 \times{10^{-5}}$ & $s^{-1}$ & Dark count	 \\[0.10cm]		
		\hline\noalign{\vskip 0.6mm}
		$e_d$ & $0.015$ & $-$ & Detection error rate 	\\[0.10cm]				
		\hline\noalign{\vskip 0.6mm}
		$\eta_d$ & $0.5$ & $-$ & Detection efficiency	\\[0.10cm]						
		\hline\noalign{\vskip 0.6mm}
		$f_e$ & $1.12$ & $-$ & Error correction efficiency	\\[0.10cm]				
		\hline\noalign{\vskip 0.6mm}
		$\eta_{filter}$ & $0.8$ & $-$ & Transmittance of the cavity filter	\\[0.10cm]				
		\hline\noalign{\vskip 0.6mm}
		$\mu_1$ & $0.002 $ & $-$ & Average photon numbers of the signal state (BB84)	\\[0.10cm]				
		\hline\noalign{\vskip 0.6mm}
		$\mu_2$ & $0.4$ & $-$ & Average photon numbers of the signal state (Decoy-state BB84)	\\[0.10cm]				
		\hline\noalign{\vskip 0.6mm}
		$\nu_2$ & $0.1$ & $-$ & Average photon numbers of the decoy state (Decoy-state BB84)	\\[0.10cm]				
		\hline\noalign{\vskip 0.6mm}
		\hline\noalign{\vskip 0.6mm}
	\end{tabular}
	\par
\end{table*}

\section*{APPENDIX C: IMPACT OF INJECTION POWER ON SECURITY OF QKD AND HACKING SUCCESS RATE}

Here, we present theoretically how an injection-locking attack affect the SKRs and QBERs of QKD systems. When a beam of light is injected into Alice's side, the frequency of Alice's laser will be shifted depending on the injection power, as is shown in FIG. 2. The main impact of our hacking strategy on the QKD systems is the increase of channel loss, which is the intrinsic loss as discussed in Appendix A.

The newly introduced transmittance of the hacking strategy $\eta _{total}$ as a function of injection power $p_{\rm{in}}$ is described by (when Alice prepares a photon in the polarization of $\left| {\rm{H}} \right\rangle \), other situations are similar):

\begin{equation}\label{eq04}
 \begin{aligned}
{\eta _{total}(p_{\rm{in}})} = \frac{1}{4}\left[ {\eta \left( {p_{\rm{H}}^{\rm{H}}} \right) + \eta \left( {p_{\rm{H}}^{\rm{V}}} \right) + \eta \left( {p_{\rm{H}}^{\rm{ + }}} \right) + \eta \left( {p_{\rm{H}}^ - } \right)} \right]
 \end{aligned}
\end{equation}
where $p_{\rm{H}}^{\rm{H}} = {p_{{\rm{in}}}}$, $p_{\rm{H}}^{\rm{V}} = 0$, and $p_{\rm{H}}^{\rm{+}} = p_{\rm{H}}^{\rm{-}} = \frac{1}{2}{p_{{\rm{in}}}}$. The subscript (superscript) of $p$ represents the polarization of the state prepared by Alice (Eve), and $\eta(p)$ is the transmittance as a function of the injection power, which can be derived from the dynamical behavior of an injected semiconductor laser presented in FIG. 2. In this calculation, we have assumed that the four polarization states generated by Alice and Eve are random. The obtained transmittance as a function of injection power is presented in FIG. 6. Bringing the transmittance into original SKRs and QBERs formulas, we obtain the SKRs and QBERs before and after the injection-locking attack, under different injection powers. Considering QKD systems using BB84 protocol, decoy-state BB84 protocol and real single photon sources (SPSs), we obtain the SKRs and QBERs shown in FIG. 5. Parameters we used in simulation are listed in TABLE IV. Meanwhile, as is shown in FIG. 6, we calculate the hacking success rate for each case, also as a function of injection power, which is described as follows:

\begin{equation}\label{eq05}
 \begin{aligned}
{\rm{hacking\; success\; rate}} = \frac{{\eta \left( {p_{\rm{H}}^{\rm{H}}} \right)}}{{\eta \left( {p_{\rm{H}}^{\rm{H}}} \right) + \eta \left( {p_{\rm{H}}^{\rm{V}}} \right) + \eta \left( {p_{\rm{H}}^{\rm{ + }}} \right) + \eta \left( {p_{\rm{H}}^ - } \right)}}
 \end{aligned}
\end{equation}

\clearpage

\end{document}